\documentclass[10pt,journal,compsoc]{IEEEtran}

\usepackage[colorlinks,urlcolor=blue,linkcolor=blue,citecolor=blue]{hyperref}

\usepackage{array}
\usepackage{mdwmath}
\usepackage{mdwtab}
\usepackage{eqparbox}
\usepackage{url}
\usepackage{balance}
\usepackage{array}
\usepackage{xcolor}
\usepackage{soul}
\usepackage[colorinlistoftodos]{todonotes}
\usepackage{longtable}
\usepackage{tabularx,booktabs}
\usepackage{rotating}
\usepackage{multirow}
\usepackage[utf8]{inputenc}
\usepackage{graphicx}
\usepackage{caption}
\usepackage{float}
\usepackage{amssymb}
\usepackage{lscape}

\begin{document}

\title{ Digital Healthcare in The Metaverse: Insights into Privacy and Security}

\author{
\IEEEauthorblockN{
\textbf{Mehdi Letafati}, \textit{Student Member, IEEE}, and
\textbf{Safa Otoum}, \textit{Member, IEEE}
}
}

\maketitle

\begin{abstract}
In this article, we study the privacy and security aspects of the metaverse in the context of digital healthcare.    
Our studies include the security aspects of data collection and communications for access to the metaverse, the privacy and security threats of employing Machine Learning and Artificial Intelligence (AI/ML) algorithms for metaverse healthcare, and the privacy of social interactions among patients in the metaverse from a human-centric perspective.  
In this article, we aim to provide new perspectives and less-investigated solutions, which are shown to be promising mechanisms in the context of wireless communications and computer science and can be considered novel solutions to be applied to healthcare metaverse services. Topics include physical layer security (PHYSec), Semantic Metaverse Communications (SMC), Differential Privacy (DP), and Adversarial Machine Learning (AML). 
As a case study, we propose \emph{distributed differential privacy} for the metaverse healthcare systems, where each virtual clinic perturbs its medical model vector to enhance privacy against malicious actors and curious servers. Through our experiments on the Breast Cancer Wisconsin Dataset (BCWD), we highlight the privacy-utility trade-off for different adjustable levels of privacy.
\end{abstract}

\maketitle

\enlargethispage{10pt}

{Metaverse} can be considered the convergence point of technologies that help facilitate immersive interactions between the physical world and virtual environments \cite{Moayad_BC}. 
One of the main applications of the metaverse is in the context of digital healthcare services, aiming to provide seamless interactions between patients and physicians for therapy sessions, medical diagnosis and/or treatment, etc. \cite{Moayad_DT}.     
For instance, patients can be treated by nurse avatars in virtual environments. This can be realized by constantly collecting the patient's medical history, symptoms, and vital signs via wearable devices and sensors.  Such e-health data is further processed and analyzed by intelligent medical agents. Finally, physicians utilize the feedback from artificial intelligence and machine learning (AI/ML)-based healthcare models when they want to provide treatments. 
	
	The metaverse's healthcare services heavily rely on extensive accessibility to collect and monitor different modalities of human life. Technological advancements in the sixth generation (6G) of wireless networks, virtual/augmented reality (VR/AR), AI/ML, digital twins, and blockchain are the key enablers for realizing metaverse healthcare \cite{Moayad_BC, Moayad_DT}.   
	Nevertheless, integrating a wide range of different technologies into the metaverse also results in the proliferation of the inherent privacy and security vulnerabilities of that technologies.    
This actually opens up different venues for privacy and security threats, which we try to address in this article.  
In Table \ref{table1}, we have provided  a summary of metaverse-enabled   services that can facilitate digital  healthcare,   
together with their corresponding use cases, privacy and security challenges, and solutions. 
	

	\begin{table*}
		\caption{Metaverse Services for Digital Healthcare and  The Corresponding Security/Privacy Challenges}
		\label{table1}
		\begin{tabular}{ |m{3.25cm}|m{4.7cm}|m{5.2cm}|m{3.0cm}|}
			\hline
			\textbf{\small  Service} &
			\textbf{\small  Description} &
			\textbf{\small  Challenges} &  \textbf{\small  Solutions}  \\
			\hline
			
			\small
			Ubiquitous data collection, and communications 
			& \small Wearables, head-mounted displays,  and other health monitoring technologies are integrated into the metaverse. 
            Such multi-modal information can be shared with health professionals to facilitate treatment decisions.  
			&   
		\small  Collection, processing, and transmission of users' sensitive information may cause privacy and security risks---AR/VR devices are prone to hacking or man-in-the-middle attacks. Efficient and reliable communication of multi-modal data is another challenge.  
		& \small PHY security \& Secure semantic communications\\
			\hline
			
			\small Distributed Medical AI 
			& \small AI/ML can be utilized in metaverse healthcare to help facilitate accurate medical diagnosis, drug development, etc.  Physicians or virtual nurses in virtual clinics and intelligent consultations utilize the feedback from the AI‑empowered medical models when treating patients. 
			& \small 
			Regulations on data privacy do not allow virtual clinics to share their medical data with other peers. Medical AI/ML  models also carry extensive information about the personal datasets, if not secured.     
			& \small Distributed DP \& AML  \\
			\hline
			
			\small Gamification \& social activities: Virtual  therapies  
			&\small  The metaverse can help facilitate the treatment of phobias, post-traumatic disorders, etc.,  through  VR therapies, providing patients with an engaging treatment experience.  
			It can also 
			provide social support via connecting patients around the world.  
			& \small   Gamification can monitor the well‑being and physical fitness of users, while this can result in leaking some private attributes, such as geographical location, habits, and lifestyles \cite{Vivek1}.    
		Besides, metaverse users can be manipulated by other individuals/avatars in a human-centric manner.
		& \small Client-level DP \& Privacy bubbles \\
			\hline
		\end{tabular}
  \vspace{-2mm}
	\end{table*}

 \section{Contributions \& State-of-the-Arts}
\begin{figure}
	\vspace{0mm}
	\includegraphics
	[width=3.25in,height=1.75in,trim={0.5in 0.7in 0.0in  1.5in},clip]{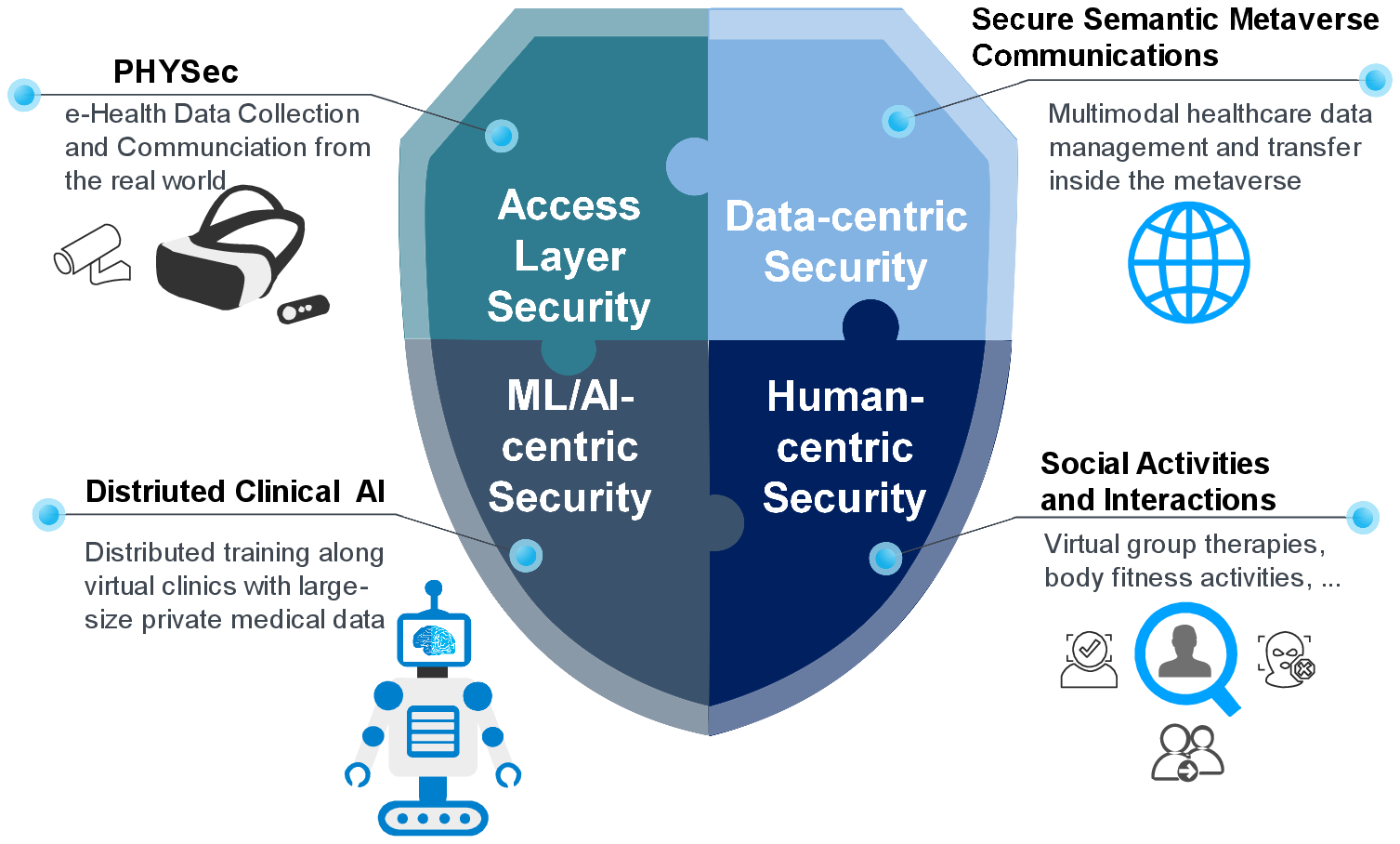}
	\vspace{-1mm}\caption{Privacy and security requirements for  metaverse healthcare.}
	\label{fig:overview}
 \vspace{-2mm}
\end{figure}

In this article, we provide \textit{guidelines and novel insights} towards enhancing the privacy and security requirements of the metaverse healthcare system.  
We should note that although a considerable number of papers address blockchain technology as an inherent means of realizing privacy- and secrecy-aware platforms \cite{Moayad_BC}, recent studies have highlighted critical vulnerabilities in the existing metaverse platforms \cite{Vivek1}.    
Therefore, instead of focusing on one specific aspect, we provide a ``holistic'' study on the privacy and security issues across different layers of the metaverse network, including the access layer, data layer, AI/ML algorithms, and human-centric challenges, as shown in Fig. \ref{fig:overview}.    

We provide new perspectives and less-investigated solutions that have been shown to be promising in the context of wireless communications and/or computer science and, hence, can be considered novel solutions to be applied to the challenges in the metaverse. 
Furthermore, as a case study, we apply the novel technique of  \emph{distributed differential privacy}\footnote{F.  Hartmann and P. Kairouz, (2023, March 2). ``\textit{Distributed differential privacy for federated learning.}'' Google Research: 
{https://ai.googleblog.com/2023/03/distributed-differential-privacy-for.html},  [Accessed Jun. 5, 2023].} to distributed learning-based healthcare systems,  hoping to shed light on further developments of distributed medical AI with formal privacy guarantees for the metaverse. 
Our globally private framework can provide an adjustable level of privacy. 
We 
evaluate our scheme on breast cancer Wisconsin dataset (BCWD)\footnote{[Online] Available: {http://archive.ics.uci.edu/ml/machine-learning-databases/breast-cancer-wisconsin/}} and highlight the privacy-utility trade-off (PUT) in terms of diagnostic accuracy for different levels of privacy.  


\subsection*{Related Papers}
To review related papers, the authors in
\cite{XAI} propose employing explainable
AI and blockchain to provide data-centric security and trust for the metaverse. The paper highlights clear examples of metaverse-enabled healthcare services. However, it lacks a “holistic” overview of different aspects of security and privacy issues that are inherently introduced by different enabling technologies (building blocks) of the metaverse. 
In \cite{BC_delay}, the authors simply focuses on the
security of AI/ML models 
Without considering any clear use case for their proposed
scheme. While proposing to exploit  blockchain technology to
safeguard the security of the learning models, the authors honestly highlight that the  inference delay of blockchain-empowered
AI/ML models make it unreasonable to integrate blockchain into the metaverse platforms as a generic solution. 
In \cite{BC_health_survey}, a systematic literature review is
provided in terms of the possible architectural mechanisms
for blockchain-based health management in the metaverse, to
fulfill interoperability and security. However, the authors
only consider the network design and architectural side of the
metaverse, while many other privacy and security challenges
(such as AI/ML security, social interactions privacy risks, etc.)
already exist in the metaverse that is not addressed in \cite{BC_health_survey}.


	\section{Safeguarding  e-Health Data Collection and Communication}\label{Sec:access} 
	Considering metaverse healthcare platforms, patients and medical doctors can use wearables with built-in sensors, AR/VR devices, e.g.,  Oculus helmets and Vive Pro headsets, and handheld controllers as the main terminals for entering the metaverse.   Such devices aim to extensively collect and analyze sensitive information of users,  such as biometric data,  speech and facial expressions, body postures,  and the surrounding environment of users.   
Therefore, lightweight and adaptive strategies are required to efficiently manage these new venues for passive and active attacks on wireless access to the metaverse. 
In terms of efficiently developing wireless access to the metaverse, 6G technology plays a pivotal role. 
Besides, 6G is the key enabler for developing novel PHY security (PHYSec) schemes as a promising approach to safeguarding the security of metaverse access, while offering formal security guarantees \cite{arxive, ICC}.  
	
\subsection{Wireless Access}
The e-health metaverse era with many peer-to-peer communications implies that we cannot rely on conventional solutions, such as the well-known  elliptic curve cryptography (ECC). Besides, the quantum computers also have the potential computation capability to break the secrecy of ECC-based algorithms \cite{6G-health}.  
Furthermore, the conventional schemes require infrastructures that cannot scale up well with the number of metaverse users and equipment. 

A promising approach to realize this paradigm shift from the conventional complexity-based solutions towards lightweight security techniques,  is PHY  secret key generation (PHY-SKG) that is  envisioned to be employed in 6G networks \cite{6G-health}. 
PHY-SKG  maintain  lightweight mechanisms with the minimum required changes at the control plane while also being able to provide  network \textit{resilience}  against adversarial attacks as well \cite{WSKG-letter, WSKG-GC}.  
PHY-SKG  can provide notable merits for access to the metaverse since it is thoroughly decentralized without relying on any infrastructure. This can substantially reduce the key generation and management time, which as a result, can make it a suitable solution  for extremely low latency applications in the metaverse, like AR/VR-based health monitoring and management.   
For establishing secure communication between wearable devices and the access point, a lightweight learning-based key agreement protocol is proposed by the authors in \cite{vtc2022}, which can shed light on future developments of device-level intelligent security for metaverse access.   


	\subsection{Multi-Modal Data Transfer: Secure Semantic Metaverse Communications (SMC)}
	\label{subsec:semantic_comm}	
	In order to provide metaverse healthcare services,  a wide  variety of human perceptions should be  reliably  exchanged among  patients, physicians, and virtual clinics. 
 Collection and communication of such an unprecedented amount of  multi-modal data can  cause critical  issues in metaverse  networks, such as network congestion. 
 To address these issues, semantic communication (SC), as a new emerging communication paradigm has shown to be capable of handling the effective transmission of multi-modal data to serve various  tasks \cite{SMC}.  
	The key idea of SC is to  simply extract the  ``relavant''   information 
	related to the users/avatars/services,  to be conveyed to the destination nodes.   
 For example, assume the semantically-extracted  data to be in the format of a  learnable embedding of a high-dimensional multi-modal data that  contains the information about  physical surrounding environments,  virtual objects/avatars,  and their  corresponding  interactions. 
	

\subsection{SMC Security: Challenge or Opportunity?!}	
	Following the SC paradigm, it can bring an inherent level of security against eavesdropping attack. In other words, 
  even if a malicious actor in the metaverse  intercepts the transmitted information, still it may  not be sufficient for obtaining  the intended  information.  
In an SMC-based metaverse network, legitimate entities  share a background knowledge, a.k.a., knowledge base (KB) for facilitating the process of  semantic encoding and decoding. Examples of this could be medical databases of authenticated virtual clinics or consultancies that are not shared by unauthorized parties.  Hence, malicious actors in the metaverse  cannot have access to the same KB, making it difficult for them  to correctly  obtain  the  semantic data.  
Moreover, a malicious agent can be interested  in a different semantic  than  that of the legitimate parties. 
		For example, a VR game captures the photos of the metaverse users, sending them to a virtual clinic in the metaverse as the legitimate receiver. While the  virtual clinic is  interested in the body fitness of its  clients,  a malicious  client  wishes to obtain the eye-tracking data.  Therefore, by employing SMC,  cameras only transmit  the semantic information that can facilitate  the task of body fitness classification. This makes it difficult (or impossible) for  malicious clients to infer their interested information, even if having perfect access to the transmitted data.  
		
	 

\begin{figure}
		\includegraphics
[width=2.9in,height=2.1in,trim={0.0in 0.0in 0.0in  0.0in},clip]{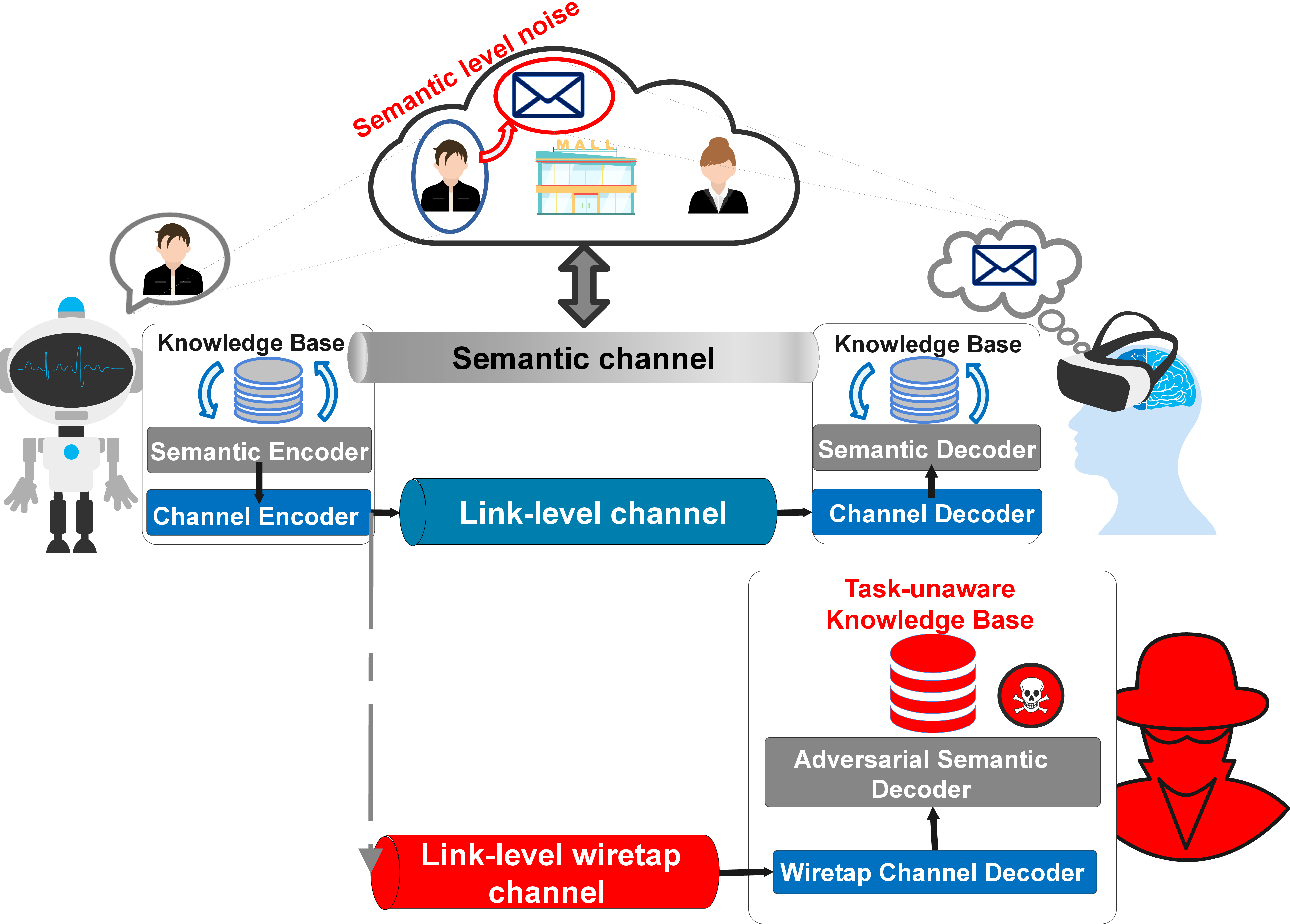}
		\vspace{2mm}\caption{SMC system security.} 
		\label{fig:SMC}
		\vspace{-3mm}
	\end{figure}

Nevertheless,  a  SMC system can be attacked by injecting  the so-called ``semantic noise'', which is the  mismatch between the original source data and the	obtained semantic information.    
For instance,  a text-based description of a patient's health status can be ``poisoned'' by substituting the synonyms  or reversing the order of some specific letters. This can result in  making the SMC system   misinterpret the semantics of data.  
To further  have an illustrative example, the abovementioned  discussions are summarized in Fig. \ref{fig:SMC}. 
To improve the robustness of SMCs against semantic noise,  adversarial training (AT)  might  be a promising approach  \cite{AML_tomg}. We highlight  that the fundamental  concepts of SMCs are still in  their infancy \cite{SMC}, let alone the security and privacy solutions. Far more in-depth studies and analysis   are required in this newly-emerged field, which  fall beyond the scope of this article.  We hope that this paper can shed light on further studies in this regard.   
Nevertheless,  the concept of AT is addressed later on in the last section. 

\section{Privacy Awareness for Social Health Activities Inside The Metaverse} \label{Sec:Data}
In a metaverse healthcare system, patients can  receive personalized care, consultancy,  and  education through interactions with other avatars and medical agents. 
Patients who  participate  in a  VR-based  group therapy session is an example of human-centric e-health activities inside the metaverse.  
Metaverse users can participate in social  activities that are purposefully created for social  health  monitoring and management. The quality of these activities rely on extracting and analyzing different personal attributes such as geographical location, habits, and life styles, making  the privacy of clients to be at risk. 
Recent studies highlight that health-related  private attributes of users, ranging  from  height and wingspan, to age and gender, can be easily compromised  during their  interactions with other  avatars and virtual environments \cite{Vivek1}. 

\begin{figure}
		\centering
		\includegraphics
		[width=2.9in,height=1.65in,trim={0.0in 0.2in 0.0in  0.0in},clip]{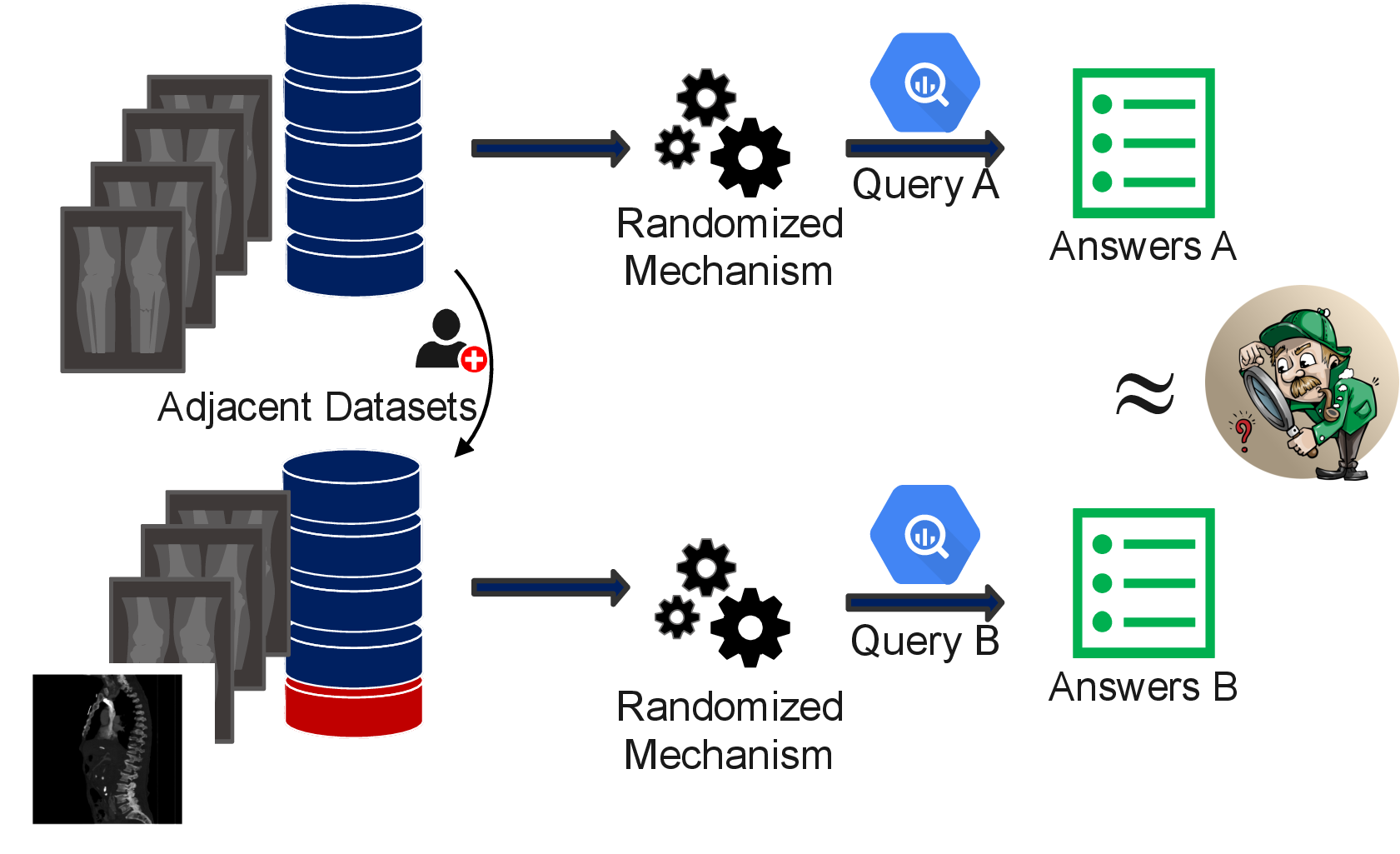}
		\vspace{-2mm}\caption{The key idea of differential privacy.}
  \label{fig:DP}
		\vspace{-4mm}
	\end{figure}

To deal with the  privacy risks  of users'  activities in the metaverse, client-level $\epsilon$-differential privacy (DP) framework can be applied, resulting in the so-called “incognito mode” metaverse.  Maintaining an adjustable level, $\epsilon$,  for privacy requirement, DP  applies randomized responses, noise, or data perturbations (depending on the format of the data attributes we aim to privatize). 
\emph{The key idea of DP is to ensure that the observable attribute profile of a metaverse user significantly overlaps with that of at least several other users, making it intractable to accurately  determine identities.} 
The key idea of DP is shown in Fig. \ref{fig:DP}. 
  


\subsection{Other Solutions for Human-Centric Privacy}
Social interactions in the metaverse inherently contain  information regarding  the clients’ preferences, habits,  mental health, etc.   
In this regard, personalized privacy preferences of users  should be effectively  addressed when developing privacy-preserving mechanisms for the metaverse \cite{Vivek2}.   

Exploiting   physical and mental  traits of users inferred during social interactions with avatars,  malicious clients 
are capable of comprehensively  understanding  users form both physical and psychological aspects.  
In other words, the metaverse users can be manipulated  by other individuals in a human-centric manner.	
To mitigate the  human-centric issues,   users should be able  to choose whether  they  wish to participate in every single virtual interaction  or not. 
Moreover, network developers should   implement privacy-enhancing  protocols, such as DP mechanism introduced in this section. 
To elaborate, users should have the option to occupy  an adjustable level of privacy ($\epsilon$ in the DP framework) that lets them dynamically manage their personal space in the virtual world, i.e., through  choosing a desired privacy configuration.
As a use-case, ``privacy bubbles'' prohibits visual access with the avatars who are outside the bubble.\footnote{{https://www.reuters.com/technology/facebook-owner-meta-adds-tool-guard-against-harassment-metaverse-2022-02-04/}{https://www.reuters.com/technology/facebook-owner-meta-adds-tool-guard-against-harassment-metaverse-2022-02-04/}} 
Users can also be provided with  secondary avatars to facilitate the process of  hiding personal attributes, traits, and preferences on demand.  
	
	 
	
Regarding the human-centric privacy and regulations,  much remains to be  investigated  in terms of  the social and personal privacy implications  of 
the metaverse, e.g.,  the  psychological aspects,  which falls beyond the scope of this article.

\section{Security and Privacy Medical AI/ML in The Metaverse} \label{Sec:AI_ML}  
	AI/ML techniques are widely  utilized in metaverse healthcare to help facilitate accurate medical diagnosis and treatment, drug development, etc., in virtual clinics and intelligent  metaverse-enabled consultations.   In this case, data privacy laws may not allow virtual
clinics or medical centres in the metaverse to share their medical data with
others. Besides,  model updates in clinical AI/ML algorithms contain  information about the medical datasets  \cite{model_inversion}.    
Currently,  most of the  intelligent  healthcare systems are based on centralized methods, which in fact  poses different  threats, such as  security and resilience  flaws and privacy leakage to metaverse network.

In this case,  federated learning (FL)-based algorithms,  as promising privacy-aware approaches,   can be considered  to be employed in the networking protocol of intelligent metaverse healthcare. 
Although FL algorithms carry  out local training without exchanging  medical data of metaverse participants,  sensitive information can still be inferred  by analyzing the model parameters  \cite{model_inversion}. As a countermeasure to privacy  leakage  and model inversions,  
a promising  approach is to employ 
differential privacy (DP) framework.  We  investigate  it  according to a case-study  on breast cancer diagnosis with  distributed DP as given in the following.

 \subsection{Case-Study: Differentially-Private Breast Cancer Diagnosis}	 
Inspired by the idea of ``incognito mode'' metaverse \cite{Vivek2}, 
we propose  \emph{distributed differential privacy} for intelligent metaverse healthcare systems.   
We utilize the idea of incognito mode to provide  \emph{global differential privacy} for distributed intelligent  metaverse healthcare systems.  
In \cite{Vivek2},  the mix-up noise is directly applied to  users' data, which is in contradiction with data cleansing processes  in data science and computer science literature.   In contrast, we apply  a randomized mechanism in the form  of artificial ``mix-up'' noise  to the federated clinical ML/AI models 
before sharing  with other peers---each of the virtual medical centers  perturbs its clinical model  parameters by adding purposefully mix-up noises before sharing them with the cloud server. Furthermore, this perturbation is adaptively fine-tuned (if needed) in the aggregation phase, to meet the privacy requirements  during global model update.  
We evaluate our scheme on breast cancer Wisconsin dataset (BCWD).\footnote{[Online] Available: {http://archive.ics.uci.edu/ml/machine-learning-databases/breast-cancer-wisconsin/}}   
The dataset contains 683 non-missing data records.  Each record is composed of 9 discrete attributes with  values in the interval $[1,10]$. There are 
458 records (i.e., 65.5\%) labeled as  ``benign'' and 241
records (i.e., 34.5\%) labeled as ``malignant.'' 
The dataset is utilized to train a federated clinical model for  classification task based on support vector machines (SVMs).  We used Python 3 to carry out our  experiments.\footnote{{https://www.python.org/downloads/release/python-3913/}}  

\begin{figure}
	\vspace{0mm}
	\centering
	\includegraphics
 [width=3.0in,height=2.2in,trim={0.0in 0.2in 0.0in  0.0in},clip]{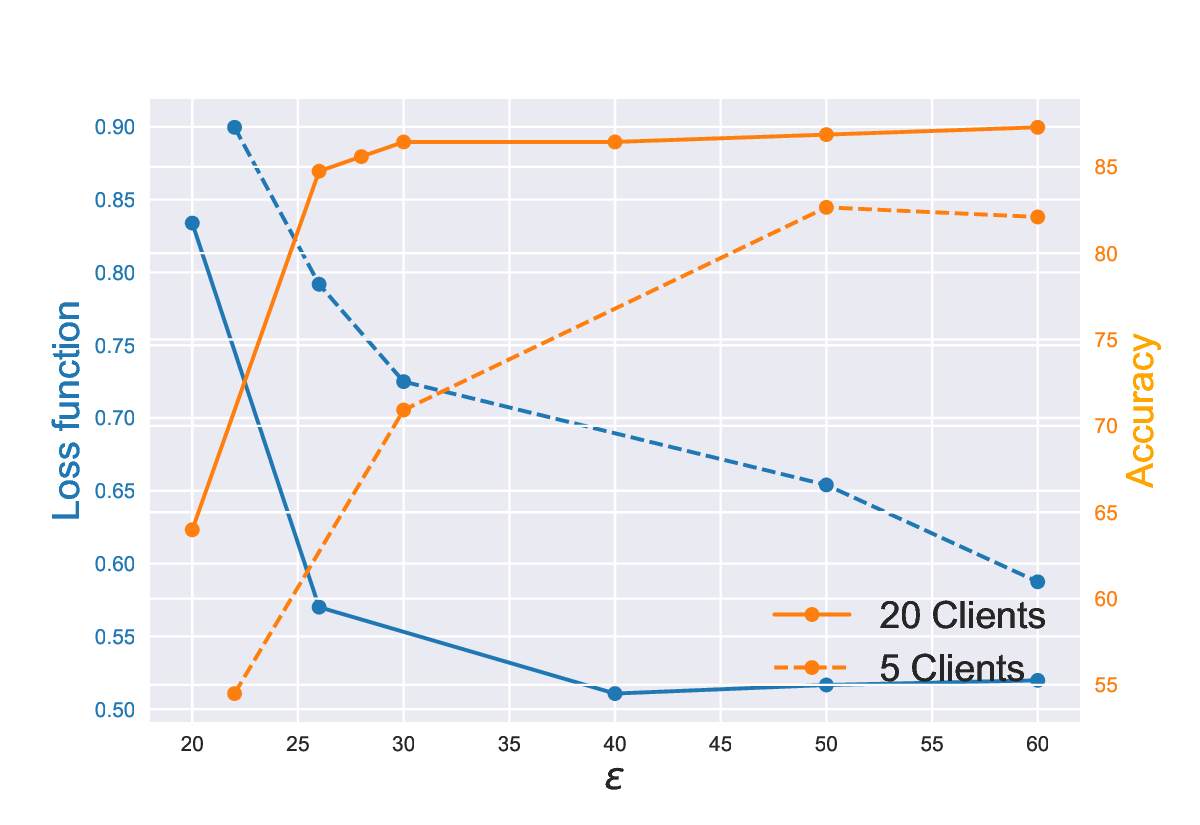}
\vspace{-3mm}
\caption{PUT: Diagnosis accuracy and loss function vs. adjustable privacy level.}
	\label{fig:trade-off}
 \vspace{-3mm}
\end{figure}

Our evaluation results are shown in Fig. \ref{fig:trade-off}.   The figure illustrates the  achieved diagnosis accuracy and the corresponding values of the loss function during learning process with different levels, $\epsilon$, of distributed DP. According to $\epsilon$-DP framework addressed in the previous section, lower $\epsilon$ values indicate more stringent privacy requirements.        
Notably,  there exist a trade-off  between the privacy protection level and the inference  performance, a.k.a privacy-utility trade-off (PUT). A better performance leads to a lower level of privacy. Nevertheless, the point is that the PUT curves start to saturate. Hence, one can for example, choose the privacy protection level $\epsilon=28$ or $\epsilon=30$ (with $20$ participating nodes) to achieve almost the same accuracy as $\epsilon=50$, while having  better privacy protection performance.    
Fig. \ref{fig:trade-off} also highlights that  increasing the number of participating clients, i.e., virtual medical clinics, higher diagnosis accuracy can be achieved, thanks to the coordination and collaboration among more distributed  medical learning parties who share their ``learning insights'' with each other.     
			
	\subsection{Medical Backdoor Attacks}      
\begin{figure}
		\centering
		\includegraphics
		[width=2.75in,height=2.8in,trim={0.0in 0.0in 0.0in  0.0in},clip]{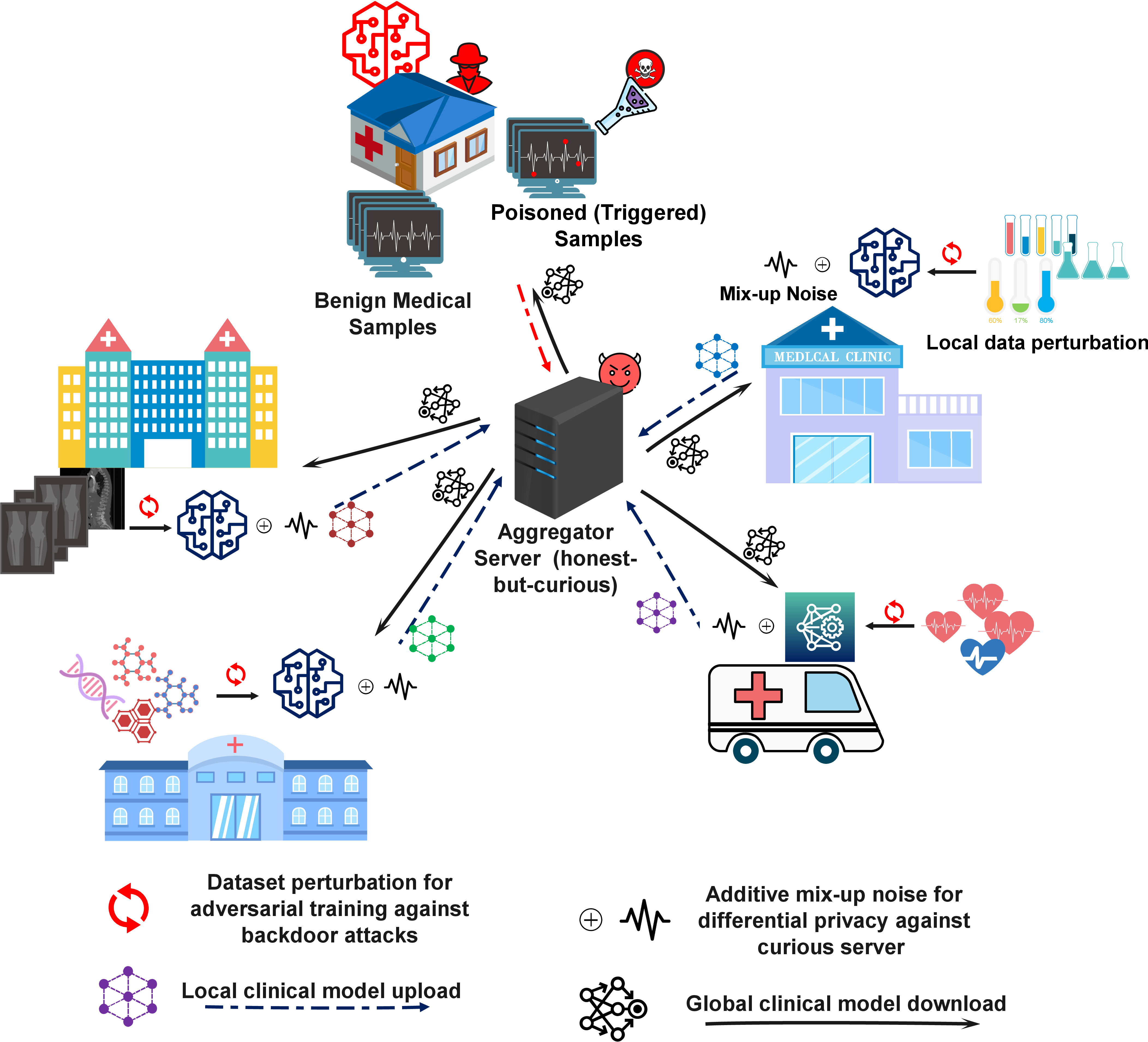}
		\vspace{2mm}\caption{Backdoor attack in distributed clinical AI and the corresponding defense mechanisms.}
		\label{fig:backdoor}
  \vspace{-4mm}
	\end{figure}

In addition to the existence of passive malicious actors who aim to infer private attributes of metaverse users,  there might exist active  adversarial participants in the metaverse, who proactively upload  ``poisoned data'' (in the format of mislabeled or manipulated medical samples) to the medical servers in order to mislead the training process of clinical AI models. This is generally  known as ``backdoor attacks''  \cite{AML_tomg}. 
 As shown in Fig. \ref{fig:backdoor}, backdoors  aim to manipulate  a subset of training data by  injecting adversarial triggers, trying to make AI/ML models 
 make incorrect (or targeted) prediction/decisions during the inference \cite{AML_tomg}.  
Backdoor attack intelligently try to  target the medical training procedure in a way that  the global  AI/ML medical  model \textit{behaves normally on untampered data, while achieves high attack success rate, i.e., incorrect decisions on poisoned data samples.}   

As an   empirically strong  defense  mechanism 
 to overcome backdoor attacks, adversarial training (AT) is  proposed. AT corresponds to the mechanism  of training a model on adversarial
	examples, with the aim of making the learning service  more resilient  to  attacks,  reducing the inference  error on	benign samples \cite{AML_tomg}.  
To summarize  the concept of private clinical AI for the  metaverse healthcare, Fig. \ref{fig:backdoor} is provided. According to this figure, with the aim of mitigating the negative effects of poisoned samples in backdoor attacks, medical data samples are first  perturbed before feeding into the clinical models.  Afterward, the clinical models (that are locally trained by each medical center in the metaverse) are combined with an artificial ``mix-up'' noise before sharing with others, in order to provide an adjustable level of differential privacy  against malicious server(s).

\section{Conclusions}
In this article, we studied   
 metaverse healthcare systems from privacy and security  perspectives.  
Our studies included  secure  data collection and communications,  private AI/ML for  distributed metaverse healthcare applications,  and the  privacy of social interactions in the metaverse from  a human-centric perspective.  
Topics included  PHYSec, secure SMC,  DP, and AT.  
  As a case-study, we proposed \emph{distributed differential privacy} for the breast cancer diagnosis in a distributed  metaverse healthcare systems.  Our evaluations on BCWD dataset  highlighted the PUT for different levels of privacy adjustment.

\section{ACKNOWLEDGMENTS}
The authors would like to thank Dr. Moayad Aloqaily for his
 indispensable comments and discussions to improve the quality of this paper.

\end{document}